\documentclass[amsmath,amssymb,prd]{revtex4-2}
\usepackage{graphicx}

\pdfoutput=1

\newcommand{\msp}{\:\!} % Space in math mode

\begin{document}

\title{Inertial Frame Dragging and Relative Rotation of ZAMOs in Axistationary Asymptotically Flat Spacetimes}

\author{Simen Braeck}
\email{simenb@oslomet.no}
\affiliation{Department of Computer Science, OsloMet – Oslo Metropolitan University, NO-0130 Oslo, Norway}

\begin{abstract}

In axistationary asymptotically flat spacetimes zero angular momentum observers (ZAMOs) define an
absolute standard of non--rotation locally, as can be verified by the absence of any Sagnac effect for these
observers. Nevertheless, we argue that on a global scale the only physically meaningful concept is that of
relative rotation. The argument is substantiated by solving Einstein's equations for an approximate thin shell
model where we keep a degree of freedom by relaxing the natural assumption of vanishing rotation at asymptotic
infinity at the outset of the analysis. The solution reveals that Einstein's equations only determine differences
in the rotation rate of ZAMOs, thereby establishing the concept of relative rotation globally. The interpretation
of rotation as relative in a global context is inherently linked to the freedom to transform between coordinate
systems rotating relative to each other, implying that an arbitrary ZAMO located at any radius may claim to be
the one who is non--rotating on a global scale and that the notion of an asymptotic Lorentz frame relative to
which one may measure absolute rotation is devoid of any meaning. The concept of rotation in Kerr spacetime
is then briefly discussed in the context of this interpretation.

\end{abstract}

\maketitle

\section{Introduction \label{sect:intro}}

The dragging of inertial frames, often called the Lense--Thirring effect, is now a well established prediction of
Einstein's general theory of relativity whereby rotating matter due to its angular momentum drags test particles or
observers with zero angular momentum (ZAMOs)~\cite{Gron1} in a co--rotating direction and cause the spin axes of gyroscopes to
precess. The effect of frame--dragging of orbits was first predicted by H. Thirring~\cite{Thirring} (1917) and J. Lense and
H. Thirring~\cite{Lense} (1918) while the closely related effect of (Schiff) frame--dragging of gyroscope axes around the
Earth was calculated by L. I. Schiff~\cite{Schiff} in 1960 and much more recently confirmed experimentally by Gravity
Probe B~\cite{GPB}.

With the continuing advancement of experimental precision and sensitivity the effects of frame-dragging may constitute
increasingly important aspects of experimental tests of general relativity as well as of other, alternative theories of
gravity in a wide range of gravitating systems. Indeed, the feasibility of detecting frame--dragging and other
gravitomagnetic effects in relation to systems such as, e.g., the planets in our solar system, the sun, supermassive
black holes and even on the laboratory scale has been investigated in several fairly recent
reports~\cite{Iorio1,Ruggiero,Iorio2,Iorio3,Iorio4,Iorio5,Iorio6,Bosi,Tartaglia}. Moreover, a test for the
Lense--Thirring effect was recently conducted even in the strong--field regime of double pulsars~\cite{Kramer}. For an
historical account of the Lense--Thirring effect, see e.g. Ref.~\cite{Pfister1}. For frequent misconceptions related to
gravitomagnetic effects, see Ref.~\cite{Costa}.

The seminal predictions by Thirring, Lense and Schiff were, however, based on approximations of slowly rotating and weak
gravitational sources of matter. D.R. Brill and J.M. Cohen~\cite{Brill} later considered an idealized model of a slowly
rotating, infinitely thin shell of matter and obtained a \emph{strong--field} solution to the dragging rate of inertial
frames by treating the geometry due to the rotating shell as a first order perturbation in the shell's angular velocity
$\omega_\mathrm{S}$ of the static Schwarzschild geometry. Then, evaluated to first order in $\omega_\mathrm{S}$, the
thin shell is spherically symmetric and spacetime in the interior of the shell is that of the flat Minkowski spacetime.
Thus, Brill and Cohen found that the angular dragging velocity $\Omega$ of the inertial frames in the exterior of the
shell steadily increases as one approaches the shell radius until it reaches a maximal value equal to a \emph{constant}
angular dragging velocity in the interior of the shell. In particular the constant angular dragging velocity in the
interior of the shell approaches the angular velocity of the shell itself as the shell mass increases and the
Schwarzschild radius approaches the shell radius. Hence they concluded that, in this limit, the inertial properties of
space in the interior of the shell do not depend on the inertial frames infinitely far away from the shell, but are
completely determined by the shell itself. This effect is often called \emph{perfect} or \emph{exact} dragging of
inertial frames. If one considers, as Brill and Cohen did, the shell of matter as an idealized model of the distant
matter in our universe, then one may establish a connection between the notion of \emph{perfect} inertial dragging and
the origin of inertia and Mach's principle. Expressed in Brill and Cohens own words: ``In this sense our result explains
why the ``fixed stars'' are indeed fixed in our inertial frame, and in this sense the result is consistent with Mach's
principle''.

Mach's principle, essentially the idea that notions of acceleration and rotation relative to an empirically unverifiable
absolute space or element are meaningless but that these quantities can be meaningfully defined only with respect to
an average motion of the total matter of the universe, and its connection with frame--dragging has been discussed in great
detail by several authors, see e.g.~\cite{Weinberg,Ciufolini,MTW,Pfister2,Schmid1,Schmid2,Gron1,Gron2,Gron3,Braeck} and references
therein. For a somewhat different viewpoint on incorporating Mach's principle in general relativity,
see also~\mbox{\cite{Khoury}}.

In a homogeneous and isotropic spacetime governed by general relativity there is perfect inertial dragging
relative to the cosmic matter \cite{Braeck}. More generally, using cosmological perturbation theory
C.~Schmid~\cite{Schmid1,Schmid2} has convincingly demonstrated perfect dragging and the validity of Mach's principle
within cosmological general relativity.

However, even if Mach's principle is demonstrably valid in a general--relativistic cosmological context, many important
solutions to Einstein's equations evidently do not share this property. In particular, this is true for the asymptotically
empty and flat solutions such as the Schwarzschild and Kerr solutions or Brill and Cohens approximate shell model, which all
approach flat Minkowski spacetime in regions far away from the localized mass distribution. These solutions are completely
devoid of any cosmic matter at great distances from the localized mass. In the far--away regions the
physical mechanism of frame dragging induced by the total matter present in these spacetimes is certainly far too weak to
account for the perfect dragging required for the inertial frames to be \emph{fully} determined by the motion of the present matter.
Invoking fictitious cosmic matter not included in Einstein's equations as external causes outside of the theory, as an
explanation for the origin of inertia, would render general relativity as a gravitational theory of spacetime and matter incomplete.
(This should not, however, be confused with the seemingly remarkable fact that the general--relativistic predictions of
frame--dragging in asymptotically flat solutions \emph{do} match the experimental measurements made relative to the fixed
stars, which could be explained by somehow merging the metric of an asymptotically flat solution with the metric determined
by the cosmic matter far away).

Asymptotic Minkowski spacetimes thus pose a challenge with regards to the interpretation of the origin of inertia in
general relativity. This rather intricate difficulty was already recognized by Einstein as early as 1917~\cite{Einstein1},
stating ``From what has now been said it will be seen that I have not succeeded in formulating boundary conditions for
spatial infinity. Nevertheless, there is still a possible way out, without resigning... For if it were possible to regard
the universe as a continuum which is \emph{finite (closed) with respect to its spatial dimensions}, we should have no need
at all of any such boundary conditions.'' The argument of incorporating Mach's principle into general relativity through
imposing restrictions on the topology of spacetime seems to have been maintained by Einstein also in his later expositions
of general relativity~\cite{Einstein2} and has been expanded upon in \cite{Ciufolini}, Chapter 5, and
in~\cite{MTW}.

It is nevertheless quite clear that somehow the local inertial frames in part are determined through
the imposed boundary conditions at infinity in asymptotically flat spacetimes. Then, it might be natural at first to assume
that this influence of the boundary conditions can be traced directly to the unique properties of the \emph{globally} empty
and nondynamical Minkowski spacetime for which there is a well-defined absolute state of non--rotation. One might, therefore,
be tempted to further infer that this nondynamical property, unaffected by the matter content of the spacetime, rather
seamlessly will be transferred to the asymptotic Minkowski spacetimes as boundary conditions ``at infinity''. In other words,
one might draw the conclusion that the ZAMOs located ``at infinity'' and ``at rest'' in asymptotic Minkowski spacetimes
correspondingly define an absolute standard of non--rotation even \emph{globally}, relative to which the orbital angular
velocity of all other ZAMOs in the spacetime is measured.

However, we note here that even if a spacetime asymptotically \emph{approaches} Minkowski spacetime, it is nowhere
\emph{exactly} flat and, in a global analysis of rotational motion, this makes the line of reasoning above questionable.
Indeed, in contrast to the conclusion drawn above, our analysis presented below indicate that only \emph{differences} in
angular velocities between ZAMOs have physical significance. This implies that we are completely free to choose any
convenient numerical reference value for the angular velocity of a ZAMO at an arbitrarily chosen radius. Only
angular velocities \emph{relative} to this arbitrary reference value are physically meaningful. As a consequence, the
absolute numerical value of the angular velocity of ZAMOs located at infinity is irrelevant, and the notion of
an absolutely non--rotating asymptotic Lorentz frame is devoid of any meaning.

Undoubtedly, in most circumstances the most practical choice for the reference value in asymptotically flat spacetimes
will be that of vanishing rotation as one approaches infinity, but fundamentally this only means that rotation of ZAMOs
is measured relative to a conveniently \emph{chosen} zero point infinitely far away (as will be clarified below).
Similarly, the importance of accounting for relative rotation implicitly appears in connection with the first law of
thermodynamics applied to Kerr--anti--de Sitter spacetimes~\mbox{\cite{Caldarelli,Gibbons}}. In Boyer-Lindquist type coordinates
for these spacetimes, the ZAMOs rotate with an angular velocity equal to the angular velocity of the black hole at the
horizon, but they also turn out to rotate with a non--zero angular velocity at asymptotic infinity in contrast to the
asymptotically flat case. In order for the first law of black hole thermodynamics to be satisfied in this case one must
use the angular velocity of the black hole measured relative to a frame that is ``non-rotating'' at infinity, i.e., it
is the relative rotation between infinity and the black hole that enters the first law. Leaving quantum effects aside,
however, the concepts of relative and absolute rotation can be discussed within general relativity independently of the
laws of black hole thermodynamics, which will be the topic of interest in this work.

\section{Inertial frame dragging in Brill and Cohen's slowly rotating shell model\label{sect:shell}}

Our purpose now is to derive an expression for the angular velocity of ZAMO's in Brill and Cohen's rotating
shell model~\cite{Brill}, but where our choice of reference point for the angular velocity is completely
arbitrary and not necessarily equal to the asymptotic boundary condition chosen at the outset in Brill and
Cohen's original work.

In their investigation of inertial frame dragging, Brill and Cohen considered an infinitely thin shell
rotating sufficiently slowly that, to first order in the shell's angular velocity $\omega_s$, the shell may be
considered spherically symmetric in shape~\cite{Hartle}. The resulting spacetime may then be treated as a small
perturbation to the spherically symmetric Schwarzschild spacetime. In isotropic coordinates, the line element for
the spacetime outside and inside the shell can then be written as
        \begin{equation}
\label{eq:lineelement}
ds^{2} = V^{2}dt^{2} - \psi^{4}{({dr^{2} + \ r^{2}{({d\vartheta^{2} + \sin ^{2}\vartheta\left( {d{\phi}
- \text{$\Omega$}\left( r \right)dt} \right)^{2}})}})}\ ,
\end{equation}
        where
        \begin{equation}
\label{eq:V}
V\left( r \right) = \left\{ {\begin{array}{c}
{\left(r - r_S\right)/\left(r + r_S\right)\quad \text{for}\ r > R} \\
{V_0\quad \text{for}\ r < R} \\
\end{array}\ ,} \right.
\end{equation}
        and
        \begin{equation}
\label{eq:psi}
\psi\left( r \right) = \left\{ \begin{array}{c}
{1 + r_S/r\quad \text{for}\ r > R} \\
{\psi_{0}\quad \text{for}\ r < R} \\
\end{array} \right..
\end{equation}
Here $\text{$\Omega$}\left( r \right)$ is the angular velocity of ZAMOs, $R$ denotes the radius of the shell, $r_S$ denotes
the shell's Schwarzschild radius, and $V_0\equiv\left(R - r_S\right)/\left(R + r_S\right)$ and $\psi_0\equiv 1 + r_S/R$ are
constants which make the components of the metric tensor continuous accross the shell. Clearly, spacetime in the interior
of the shell is then manifestly flat Minkowski spacetime expressed in conveniently scaled coordinates.

If, as Brill and Cohen did in their original work, we now impose the asymptotic boundary condition
$\displaystyle \lim_{r\rightarrow \infty} \text{$\Omega$}\left( r \right) = 0$ at the outset, then the line element above
approaches that of Minkowski spacetime in standard ``non--rotating'' spherical coordinates. However, in so doing one may
in the final result end up with the wrong impression that the inertial frames at infinity somehow, without any choice of
freedom, single out a global standard of non--rotation. For this reason we shall not impose any boundary conditions on
$\text{$\Omega$}\left( r \right)$ at this stage in the derivation, but instead keep this degree of freedom temporarily
until the boundary condition naturally is to be determined at a later stage.

We may now use Einstein's equations in combination with the line element given above in order to find the explicit
expression for $\text{$\Omega$}\left( r \right)$. A detailed step--by--step rederivation of Brill and Cohen's original result
with the restriction $\displaystyle \lim_{r\rightarrow \infty} \text{$\Omega$}\left( r \right) = 0$ at the outset has already
been given in Ref.~\cite{Braeck}. The derivation for the more general case with no such restriction on
$\text{$\Omega$}\left( r \right)$ at the outset is essentially identical to the one presented in Ref.~\cite{Braeck}.
Hence we shall not repeat the derivation in full detail here, but only give an outline of that derivation while we keep
track of where the modifications to Einstein's equations, $\displaystyle G_{\alpha\beta}=8\pi GT_{\alpha\beta}$, occur
underway.

For the present purpose Einstein's equations are most easily solved by using the Cartan formalism~\cite{Hervik}.
In this context a useful set of orthonormal basis one--forms are given by
\begin{equation}
\label{eq:OB1}
\omega^{0} = Vdt,\ \ \omega^{1} = \psi^{2}dr,\ \ \omega^{2} = r\psi^{2}d\vartheta,\ \
\omega^{3} = r\psi^{2}\sin \vartheta\left( {d{\phi} - \text{$\Omega$}dt} \right)\,.
\end{equation}
From Cartan's structural equations we then find for the non--zero components of the Einstein tensor,
        \begin{equation}
\label{eq:ET00}
G^{00} = \frac{4r_{S}}{r^{2}\psi^{5}}\delta\left( {r - R} \right)\ ,
\end{equation}
\begin{equation}
\label{eq:ET22}
G^{22} = G^{33} = \frac{r_{S}}{2r\psi V}G^{00}\ ,
\end{equation}
\begin{equation}
\label{eq:ET03}
G^{03} = - \frac{\sin \vartheta}{2r^{3}\psi^{8}}\left( \frac{r^{4}\psi^{6}\text{$\Omega$}^{\prime}}{V} \right)^{\prime}\ .
\end{equation}
Here $\delta\left( r \right) $ is the Dirac delta function, and $V$ and $\psi$ are given in Equations~(\ref{eq:V}) and
(\ref{eq:psi}), respectively. For the diagonal components of the stress--energy tensor $T^{\mu\nu}$, Einstein's field
equations now immediately yield
  \vspace{12pt}
      \begin{equation}
\label{eq:SET00}
{\rho}\equiv T^{00} = \frac{G^{00}}{8\pi} = \frac{r_{S}r^{3}}{2\pi\left( {r + r_{S}} \right)^{5}}\delta\left( {r - R} \right)\ ,
\end{equation}
\begin{equation}
\label{eq:SET}
T^{33} = T^{22} = \frac{G^{22}}{8\pi} = \frac{r_{S}}{2\left( {r - r_{S}} \right)}{\rho}\ .
\end{equation}
Here ${\rho} $ denotes the mass density of the shell in the rest frame of an element of the shell.

To proceed, we next consider the Einstein equation containing the nondiagonal components,
$\displaystyle G^{03}=8\pi GT^{03}$. Because spacetime is empty both in the interior ($r < R $) and
exterior ($r > R $) of the shell, we have that $T^{03} = 0$ and this equation reduces to
\begin{equation}
\label{eq:nd}
\left( \frac{r^{4}\psi^{6}\text{$\Omega$}^{\prime}}{V} \right)^{\prime} = 0\ r \neq R\ .
\end{equation}
Thus, we find
\begin{equation}
\label{eq:omprime}
\text{$\Omega$}_{\pm}^{\prime} = \frac{K_{\pm}V}{{r^4\psi^{6}}}\ .
\end{equation}
In this expression $K_{-} $ and $K_{+} $ are constants of integration in the two regions $r < R $ and $r > R $, respectively.
For $r < R $, we have $\psi = \psi_{0} $ and $V = V_{0} $. Hence, we obtain
\begin{equation}
\label{eq:intomega}
\text{$\Omega$}_{-} = - \frac{K_{-}V_{0}}{3r^{3}\text{$\psi$}_{0}^{6}} + \text{$\Omega$}_{B}\ ,
\end{equation}
where $\text{$\Omega$}_{B} $ is another constant of integration yet to be determined. In the shell's interior spacetime is
assumed to be flat. We therefore require a regular solution as $r\rightarrow 0$, from which it follows that $K_{-} = 0$ and
$\text{$\Omega$}_{-}=\text{$\Omega$}_{B}$ . For $r > R$ we may integrate equation~(\ref{eq:omprime}) to give
\begin{equation}
\label{eq:extomega}
\text{$\Omega$}_{+}\left( r \right) - \text{$\Omega$}_{+}\left( r_Q \right) =
K_{+}\int_{r_Q}^{r}\frac{r-r_S}{r^5\psi\left(r\right)^7}\msp dr
=-\frac{K_{+}}{3}\left[\frac{1}{r^3\psi\left(r\right)^6} - \frac{1}{r_Q^3\psi\left(r_Q\right)^6}\right]\ .
\end{equation}
Here $r_Q$ is an arbitrarily chosen reference radius. It is important to note here that Einstein's equation determines
only the \emph{difference}, $\text{$\Omega$}_{+}\left( r \right) - \text{$\Omega$}_{+}\left( r_Q \right)$,
in angular velocity between ZAMOs located at two different radii. Moreover, this difference is completely independent of the
numerical value of $\text{$\Omega$}_{+}\left( r_Q \right)$. Thus we are free to choose \emph{any} convenient reference value
$\text{$\Omega$}_Q\equiv\text{$\Omega$}_{+}\left( r_Q \right)$ at the reference point $r_Q$. In essence, the choice of the
reference point and reference value for the angular velocity here is analogous to the arbitrary choice of a reference point for
potential energy in classical mechanics or to the arbitrary choice of a specific inertial frame for measuring velocities.
In practice this always allow us to conveniently define a new angular velocity function
$\text{$\Omega$}_{\mathrm{rel}}\left( r \right)\equiv
\text{$\Omega$}_{+}\left( r \right) - \text{$\Omega$}_{Q}$ which then describes the local rotation of inertial
frames \emph{relative} to the \emph{arbitrarily chosen} local rotation $\text{$\Omega$}_Q$ of inertial frames located at the
arbitrarily chosen reference radius $r_Q$. Equivalently, one may instead simply declare the angular velocity $\text{$\Omega$}_Q$
of the ZAMOs located at the radius $r_Q$ to be zero. The angular velocity $\text{$\Omega$}_{+}\left( r \right)$ then describes
the rotation rate of the ZAMOs \emph{relative} to the arbitrarily chosen zero rotation rate of ZAMOs located at $r_Q$.
Finally, this analysis makes it clear that the angular velocity of ZAMOs located at asymptotic infinity play no particular role
in determining a reference value for the angular velocity in this spacetime.

The constant $K_{+}$ may now be determined by requiring the metric to be continuous accross the shell,
$\text{$\Omega$}_{-}\left( R \right) = \text{$\Omega$}_{B} = \text{$\Omega$}_{+}\left( R \right)$,
giving
\begin{equation}
\label{eq:Kplus}
K_{+} = - \frac{3\left(\text{$\Omega$}_{B} -\text{$\Omega$}_{Q}\right)}
{\frac{1}{r^3\psi\left(r\right)^6} - \frac{1}{r_Q^3\psi\left(r_Q\right)^6}}\,.
\end{equation}
Thus, the angular velocity in the two regions can be expressed as
\begin{equation}
\label{eq:omegap}
\text{$\Omega$}\left( r \right) = \left\{ {\begin{array}{c}
\text{$\Omega$}_{Q} + \frac{\left[\frac{1}{r^3\psi\left(r\right)^6} - \frac{1}{r_Q^3\psi\left(r_Q\right)^6}\right]
\left(\text{$\Omega$}_{B} -\text{$\Omega$}_{Q}\right)}
{\frac{1}{R^3\psi_0^6} - \frac{1}{r_Q^3\psi\left(r_Q\right)^6}}\ \text{for}\ r > R \\
{\ \text{$\Omega$}_{B}\ \text{for}\ r < R} \\
\end{array}\ .} \right.
\end{equation}
Combining this expression with equation~(\ref{eq:ET03}), the nondiagonal component of the Einstein tensor can
be written as
\begin{equation}
\label{eq:ET03B}
G^{03} = \frac{3\left(\text{$\Omega$}_{B} -\text{$\Omega$}_{Q}\right)
\left( R\psi_{0}^{2} \right)^{3}\sin \upsilon}
{1-\left(\frac{r_Q\left(R+r_S\right)^2}{R\left(r_Q+r_S\right)^2}\right)^3}\delta\left( {r - R} \right)\ .
\end{equation}

Our next task is to determine the constant $\text{$\Omega$}_{B}$ entering the expression above. This we may accomplish
by once again integrating Einstein's equation, $\displaystyle G^{03}=8\pi GT^{03}$, but this time over a region
crossing the shell radius $R$. Accordingly, we must first consider the stress--energy tensor of the shell. From the
requirement that the momentum densities $T^{i0}$ must vanish in the rest frames of the matter comprising the shell,
Brill and Cohen~\cite{Brill} deduced that this stress--energy tensor should have the form
\begin{equation}
\label{eq:SETshell}
T^{\mu\nu} = {\rho} u^{\mu}u^{\nu} + \sum\limits_{i,j = 1}^{3}t^{ij}v_{(i)}^{\mu}v_{j}^{\nu}\ ,
\end{equation}
where as before ${\rho} $ denotes the mass density in the rest frame of the shell, $u^{\mu} $ are the components of the
four-velocity of a given element of the shell and $v_{(i)}^{\mu} $ are the components of three spatial orthonormal vectors
spanning the hypersurface orthogonal to $u^{\mu}$. We shall here also assume that this form of the stress--energy tensor
is adequate to first order in angular velocities.

We now proceed to find the components $T^{\mu\nu}$ of the shell. Let each element of the shell rotate with a given angular
velocity $d\phi/dt=\omega_s$ in the isotropic coordinates. Using that $dr = d\vartheta = 0$ for the element in the line
element~(\ref{eq:lineelement}), the components of the four--velocity in the coordinate basis are calculated as
\begin{equation}
\label{eq:FVcoord}
{\widetilde{u}}^{0} = \frac{dt}{d\tau} = \frac{1}{V_0}\left( {1 - \sigma^{2}} \right)^{- 1/2}\ ,
\end{equation}
\begin{equation}
{\widetilde{u}}^{1} = {\widetilde{u}}^{2} = 0\ ,
\end{equation}
\begin{equation}
{\widetilde{u}}^{3} = \omega_{s}{\widetilde{u}}^{0}\ ,
\end{equation}
where we have introduced the quantity
\begin{equation}
\label{eq:sigma}
\sigma = \frac{R\psi_0^{2}\sin \vartheta\left( {\omega_{s} - \text{$\Omega_B$}} \right)}{V_0}\ .
\end{equation}
From the relations~(\ref{eq:OB1}) between the orthonormal basis and the coordinate basis, the components
of the four--velocity in the orthonormal basis are obtained as
\begin{equation}
\label{eq:FVortho0}
u^{0} = \left( {1 - \sigma^{2}} \right)^{- 1/2}\ ,
\end{equation}
\begin{equation}
\label{eq:FVortho1}
u^{1} = u^{2} = 0\ ,
\end{equation}
\begin{equation}
\label{eq:FVortho3}
u^{3} = \sigma\left( {1 - \sigma^{2}} \right)^{- 1/2}\ .
\end{equation}

By choosing the components of the three spatial vectors in the orthonormal basis as
\begin{equation}
\label{eq:tri1}
v_{(1)}^{\mu} = \left( {0,\ 1,\ 0,\ 0} \right)\ ,
\end{equation}
\begin{equation}
\label{eq:tri2}
v_{(2)}^{\mu} = \left( {0,\ 0,\ 1,\ 0} \right)\ ,
\end{equation}
\begin{equation}
\label{eq:tri3}
v_{(3)}^{\mu} = \left( {\sigma,\ 0,\ 0,\ 1} \right)\left( {1 - \sigma^{2}} \right)^{- 1/2}\ ,
\end{equation}
we ensure that they have unit lengths and are orthogonal to each other, as well as being orthogonal
to the four--velocity. Using these expressions for $u^{\mu}$ and $v_{(i)}^{\mu}$ in Equation~(\ref{eq:SETshell})
in combination with the results already obtained in~(\ref{eq:SET00}) and (\ref{eq:SET}), the components $T^{\mu\nu}$
are, to first order in $\omega_s-\text{$\Omega$}_{B}$, calculated to be
\begin{equation}
\label{eq:SEap00}
T^{00} = \rho\ ,
\end{equation}
\begin{equation}
\label{eq:SEap22}
T^{22} = \ \frac{\rho r_{S}}{2\left( {R - r_{S}} \right)} = \ t^{22},
\end{equation}
\begin{equation}
\label{eq:SEap33}
T^{33} = \ T^{22} = \ t^{33},
\end{equation}
\begin{equation}
\label{eq:SEap03}
T^{03} =  \ \left( \rho + t^{33} \right)\sigma = \ \frac{\rho\sigma\left(2R-r_S\right)}{2\left( {R - r_{S}} \right)} \ .
\end{equation}
If we now integrate Einstein’s field equation, $G^{03} = 8\pi T^{03}$, across the shell's radius $R$ we obtain the
equation
\begin{equation}
\label{eq:EFE03}
\frac{3\left(\text{$\Omega$}_{B} - \text{$\Omega$}_{Q}\right)R^{2}\sin \vartheta}{2\left( {R + r_{S}} \right)^{2}
\left[1-\left(\frac{r_Q\left(R+r_S\right)^2}{R\left(r_Q+r_S\right)^2}\right)^3\right]}
= \frac{2r_{S}\sin \vartheta \left( 2R-r_S \right)R^{2}\left( {\omega_{s} -
\text{$\Omega$}_{B}} \right)}{\left( {R + r_{S}} \right)^{2} \left( {R - r_{S}} \right)^2}\ ,
\end{equation}
which may be readily solved to give
\begin{equation}
\label{eq:OmegaB}
\text{$\Omega$}_{B} = \dfrac{\omega_{s} + \dfrac{3\left( {R - r_{S}} \right)^2}
{4r_S\left(2R-r_S\right)\left[1-\left(\frac{r_Q\left(R+r_S\right)^2}{R\left(r_Q+r_S\right)^2}\right)^3\right]}\,\text{$\Omega_Q$}}
{1 + \dfrac{3\left( {R - r_{S}} \right)^2}
{4r_S\left(2R-r_S\right)\left[1-\left(\frac{r_Q\left(R+r_S\right)^2}{R\left(r_Q+r_S\right)^2}\right)^3\right]}}\ .
\end{equation}
We now substitute this result for $\text{$\Omega$}_{B}$ in Equation~(\ref{eq:omegap}) to obtain our final result for
the angular velocity of the inertial frames as
\begin{equation}
\label{eq:finalomega}
\text{$\Omega$}\left( r \right) =  \left\{ {\begin{array}{l}
\text{$\Omega_Q$}
+\dfrac{g\left(r,r_Q\right)\left(\omega_s-\text{$\Omega_Q$}\right)}
{g\left(R,r_Q\right)\left(1 + \dfrac{3\left( {R - r_{S}} \right)^2}
{4r_S\left(2R-r_S\right)\,h\left(R,r_Q\right)}\right)}\ \ \ \ \
 \text{for}\ r > R \\ \\
\text{$\Omega_Q$} + \dfrac{\omega_{s} - \text{$\Omega_Q$}}
{1 + \dfrac{3\left( {R - r_{S}} \right)^2}{4r_S\left(2R-r_S\right)\,h\left(R,r_Q\right)}}
\ \ \ \ \ \text{for}\ r < R \\
\end{array}\ ,} \right.
\end{equation}
where
\begin{equation}
\label{eq:funcomega1}
g\left(r,r_Q\right) = \left(\dfrac{r}{\left( r + r_S\right)^2}\right)^3
- \left(\dfrac{r_Q}{\left( r_Q + r_S\right)^2}\right)^3\ ,
\end{equation}
\begin{equation}
\label{eq:funcomega2}
h\left(R,r_Q\right) = 1-\left(\frac{r_Q\left(R+r_S\right)^2}{R\left(r_Q+r_S\right)^2}\right)^3\ .
\end{equation}
We note here that $\text{$\Omega$}\left( r \right)$ depends only on the \emph{difference}
$\omega_s-\text{$\Omega_Q$}$, i.e., only on the angular velocity of the shell \emph{relative}
to the arbitrary reference angular velocity of the inertial frames located at $r_Q$, and not on the
numerical value of $\omega_s$ alone. Accordingly, if the difference $\omega_s-\text{$\Omega_Q$}$ is kept fixed,
any difference $\text{$\Omega$}\left( r_2 \right) - \text{$\Omega$}\left( r_1 \right)$ in the angular
velocity of the inertial frames at different radii is independent of the arbitrary reference value
$\text{$\Omega_Q$}$. This reflects the fact that the numerical value of $\text{$\Omega$}\left( r \right)$ is
insignificant; only \emph{relative} angular velocities have real physical significance. The difference in
angular velocities of inertial frames may of course vanish, as in the interior of the shell, in which case
the concepts of relative and absolute rotation become indistinguishable.

The explicit appearance of the term $\text{$\Omega_Q$}$ in~(\ref{eq:finalomega}) can be understood as follows. If
the angular velocity of the shell approaches the reference value, i.e., $\omega_s\rightarrow \text{$\Omega_Q$}$,
the angular velocity $\text{$\Omega$}\left( r \right)$ of the ZAMOs approaches the constant reference value
$\text{$\Omega_Q$}$ both in the interior \emph{and} exterior regions of the shell. This limit therefore describes
a gradual transition from a situation in which the ZAMOs in the exterior region rotate relative to each other to a
situation where all the ZAMOs eventually end up perfectly co--rotating with an angular velocity $\text{$\Omega_Q$}$
relative to the coordinate system everywhere throughout spacetime. The resulting spacetime is accordingly that of
Minkowski and Schwarzchild spacetime in the interior and exterior of the shell, respectively, as described in a
coordinate system rotating relative to the collection of co--rotating ZAMOs. Thus, since the relative angular
velocities of the ZAMOs completely vanish in this limit, we evidently recover a spacetime with the ``apparent''
property of absolute rotation everywhere analogous to that of global Minkowski spacetime. We emphasize that
the transition occuring here is completely independent of any reference to non--rotating ZAMOs at infinity or
non--rotating asymptotic Lorentz frame.

To gain some further insight, consider now for simplicity the case $\omega_s = \text{$\Omega_Q$}$ in the interior region $r<R$ of the shell. Then again $\text{$\Omega$}\left( r \right) = \text{$\Omega_Q$}$, and the line element
in~(\ref{eq:lineelement}) simplifies to that of
\begin{equation}
\label{eq:intlineelement}
ds^{2} = \left( 1-{r^{\prime}}^2\sin ^{2}\vartheta\, \text{$\Omega^{\prime}_Q$}^2 \right)d{t^{\prime}}^{2}
- d{r^{\prime}}^2 - {r^{\prime}}^2d\vartheta^2 - {r^{\prime}}^2\sin ^{2}\vartheta\, d{\phi}^2
+ 2{r^{\prime}}^2\sin ^{2}\vartheta\, \text{$\Omega^{\prime}_Q$}\,d{t^{\prime}}^{2}d{\phi}^2\ ,
\end{equation}
where we have introduced the conveniently rescaled coordinates $r^{\prime}=\psi_0^2\, r$, $t^{\prime}=V_0\, t$,
and $\text{$\Omega^{\prime}_Q$}=d{\phi}/dt^{\prime}=\text{$\Omega_Q$}/V_0$. We recognize the line element above as
that of Minkowski spacetime described in a spherical coordinate system rotating with constant angular velocity
$-\text{$\Omega^{\prime}_Q$}$ relative to a standard inertial reference frame in which the ZAMOs do not rotate.
Unless $\text{$\Omega^{\prime}_Q$}=0$, it is clear that inertial effects will appear in the interior of the shell
in this rotating coordinate system. More generally, when $\omega_s \neq \text{$\Omega_Q$}$, these inertial effects
will increase if $\omega_s > \text{$\Omega_Q$}$ due to the second term in Equation~(\ref{eq:finalomega})
for $r<R$ which accounts for additional rotational effects caused by increased rotation of the shell relative to
ZAMOs located at the reference radius $r_Q$, implying also an increased rotation rate of the shell relative to the
coordinate system. Alternatively, keeping the parameters $\omega_s$ and $\text{$\Omega_Q$}$ fixed, this second
term also implies that these inertial effects depend upon the choice of the reference point $r_Q$. For instance,
as $r_Q\rightarrow \infty$ the function $h\left(R,r_Q\right)\rightarrow 1$, yielding a rotation rate greater than
$\text{$\Omega_Q$}$. On the other hand, for $r_Q\rightarrow R$, $h\left(R,r_Q\right)\rightarrow 0$ such that the
second term vanishes and hence the rotation rate approaches $\text{$\Omega_Q$}$. This dependence on $r_Q$ may at
first seem suspect, but is effectively caused by changes in the relative rotation between ZAMOs in the exterior
region of the shell which impacts the rotation rate in the interior.

\begin{figure}
\begin{center}
\includegraphics[width=14.0cm,trim={0.0cm 8.2cm 0.5cm 9.0cm},clip]{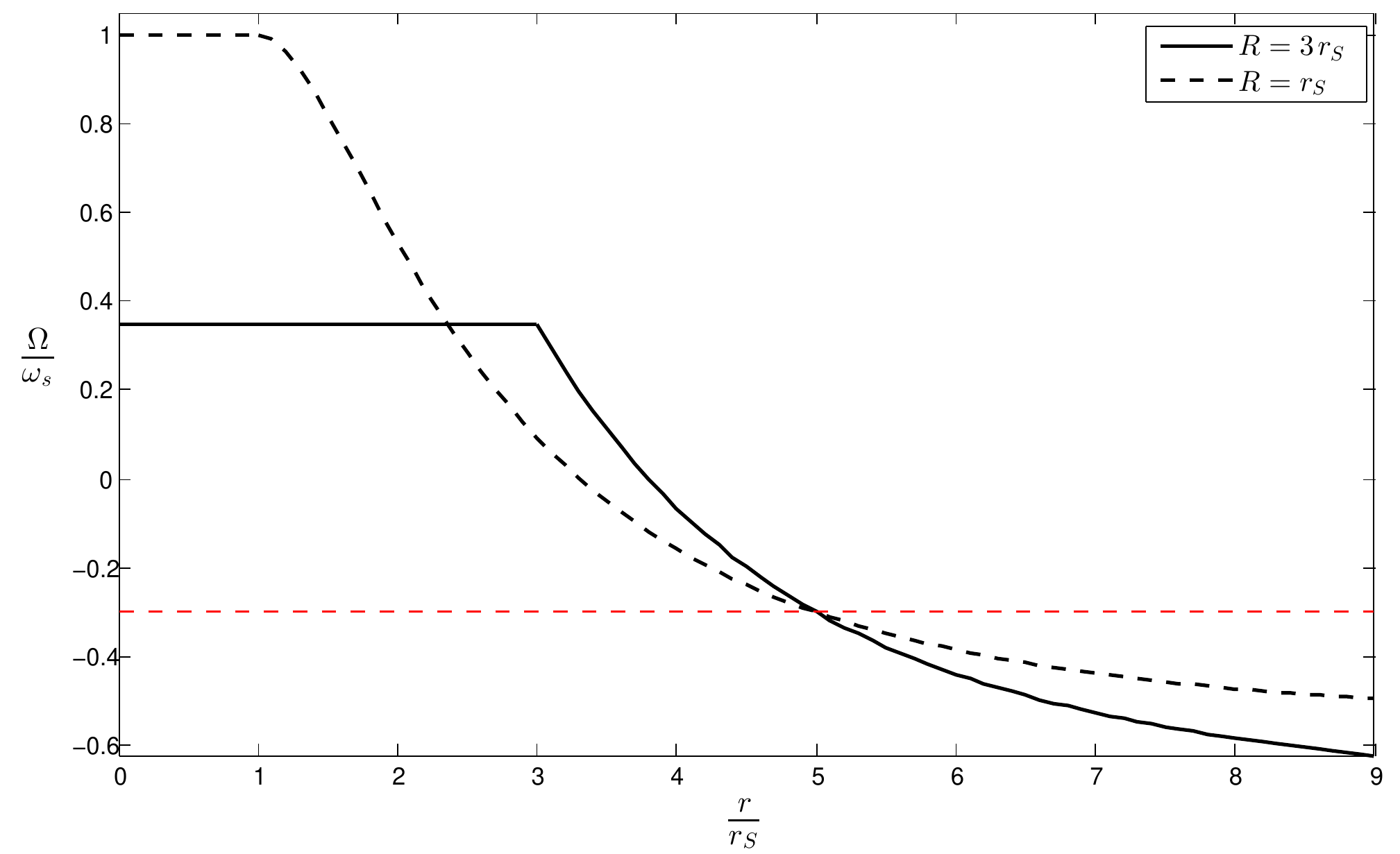}
\caption{Plots of the angular velocity $\text{$\Omega$}\left( r \right)$ of ZAMOs as given by the expression in
Equation~(\ref{eq:finalomega}) with $\text{$\Omega$}_Q=-0.3\,\omega_s$ and $r_Q=5\,r_S$. Note that
$\text{$\Omega$}\left( r \right)$ is normalized by $\omega_s$. The solid curve shows the
result for a shell radius $R=3\,r_S$. The black dashed curve shows the result for a shell radius equal to its
Schwarzschild radius, $R=r_S$. This corresponds to the situation of ``perfect inertial dragging'' where the inertial
frames in the interior of the shell rotate with the same angular velocity as the shell, independently of the angular
velocity of the inertial frames located ``at infinity''. The red dashed curve marks the chosen value for
$\text{$\Omega$}_Q$ (normalized by $\omega_s$).}
\label{fig:gensol}
\end{center}
\end{figure}

Figure~\ref{fig:gensol} illustrates two examples of the angular velocity $\text{$\Omega$}\left( r \right)$
in~(\ref{eq:finalomega}) for fixed parameter values $\text{$\Omega$}_Q=-0.3\,\omega_s$ and $r_Q=5\,r_S$,
and for two different choices, $R=3\,r_S$ and $R=r_S$, for the shell radius.
As can be seen, when the shell radius is larger than the Schwarzschild radius, the inertial frames are only partially
dragged around with the shell's rotation. As the shell radius approaches its Schwarzschild radius, however,
the inertial frames in the interior of the shell rotate with the same angular velocity $\omega_s$ as the
shell, independently of the angular velocity of the inertial frames located ``at asymptotic infinity''. This is the
phenomenon of ``perfect inertial dragging'' first discovered by Brill and Cohen~\cite{Brill}. Yet, in Brill and Cohen's
original calculation of $\text{$\Omega$}\left( r \right)$ the inertial frames located ``at infinity'' were
non--rotating. In contrast, with our choice above for the reference value $\text{$\Omega$}_Q$, the inertial frames
located ``at infinity'' in our case rotate with a negative angular velocity; in other words, they are not at rest.

\begin{figure}
\begin{center}
\includegraphics[width=14.0cm,trim={0.0cm 8.2cm 0.5cm 9.0cm},clip]{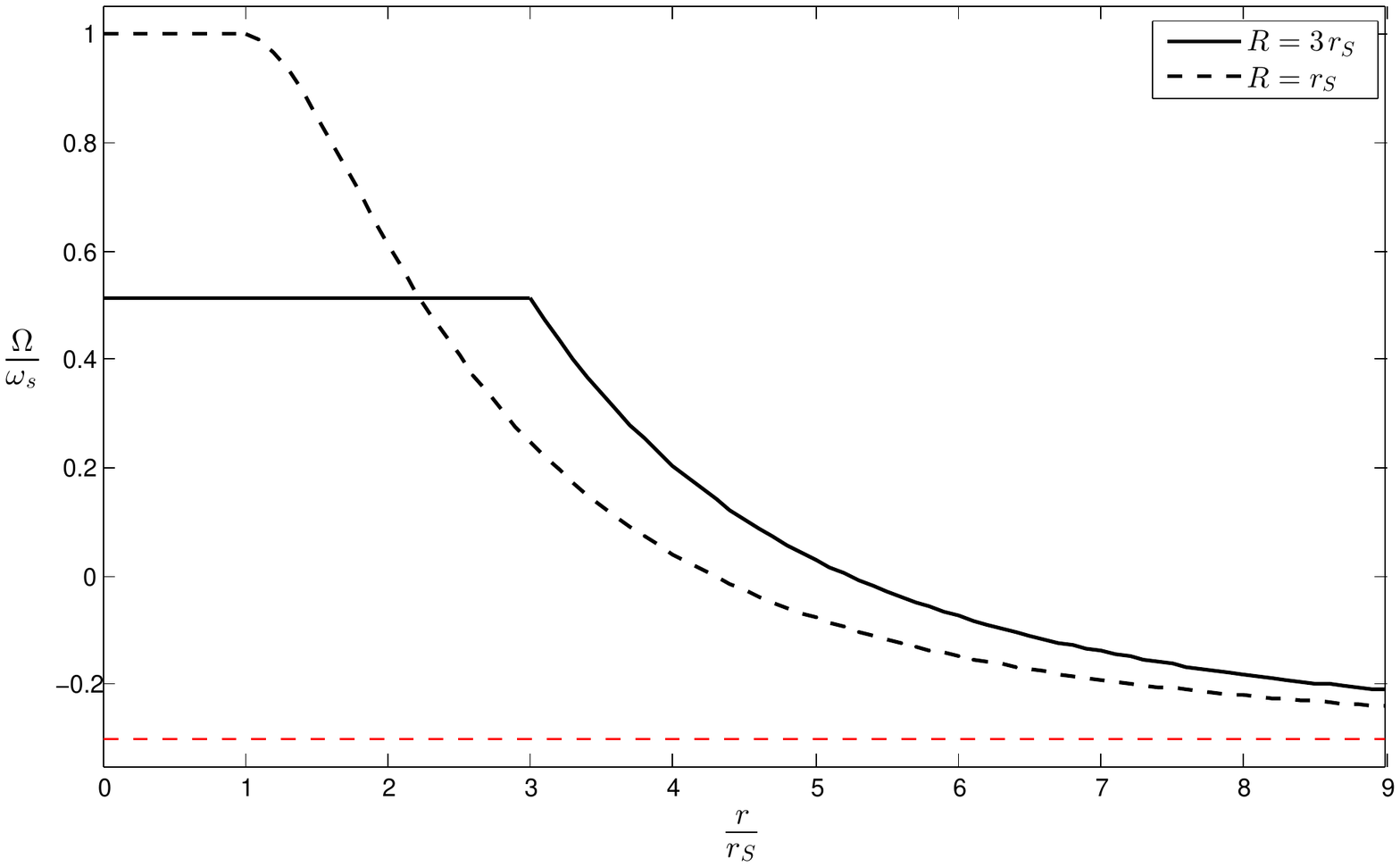}
%\vspace{-0.1cm}
\caption{Same parameter values as in Figure~\ref{fig:gensol} except that here $r_Q\rightarrow\infty$.}
\label{fig:gensolinf}
\end{center}
\end{figure}

To further compare our result to the original result of Brill and Cohen, it is instructive to consider the case
for which $r_Q\rightarrow\infty$, as shown in Figure~\ref{fig:gensolinf}.
Then the ZAMOs located at asymptotic infinity
rotate with the angular velocity $\text{$\Omega$}_Q$. The original result of Brill and Cohen is now recovered
by making the convenient, but very special choice $\text{$\Omega$}_Q=0$, for which Equation~(\ref{eq:finalomega})
simplifies to
\begin{equation}
\label{eq:BComega}
\text{$\Omega$}\left( r \right) =  \left\{ {\begin{array}{l}
\dfrac{\left(\dfrac{r\left( R+r_S \right)^2}{R\left( r + r_S\right)^2}\right)^3\omega_s}
{1 + \dfrac{3\left( {R - r_{S}} \right)^2}
{4r_S\left(2R-r_S\right)}}\ \ \ \ \
 \text{for}\ r > R \\ \\
\dfrac{\omega_{s}}
{1 + \dfrac{3\left( {R - r_{S}} \right)^2}{4r_S\left(2R-r_S\right)}}
\ \ \ \ \ \text{for}\ r < R \\
\end{array}\ .} \right.
\end{equation}

Note that Equation~(\ref{eq:finalomega}) can be obtained from Equation~(\ref{eq:BComega}) by transforming to a coordinate
system $\left({\widetilde{t},\widetilde{r},\widetilde{\vartheta},\widetilde{{\varphi}}}\right)$ rotating relative to the
first one with an angular velocity
$\omega = \text{$\Omega$}\left( r_Q \right)-\widetilde{\text{$\Omega$}}\left( r_Q \right)$:
\begin{equation}
\label{eq:transf}
\widetilde{t}=t, \ \ \ \widetilde{r}=r, \ \ \ \widetilde{\vartheta}=\vartheta,
\ \ \ \widetilde{{\varphi}}=\varphi - \left(\text{$\Omega$}\left( r_Q \right)-\widetilde{\text{$\Omega$}}_Q\right)t\,,
\end{equation}
where $\widetilde{\text{$\Omega$}}_Q\equiv \widetilde{\text{$\Omega$}}\left( r_Q \right)$ is the arbitrarily chosen
reference value for the angular velocity of the ZAMOs located at the reference radius $r_Q$ in the rotating coordinate
system. The angular velocity of the rotating shell in the rotating coordinate system is then
$\widetilde{\omega}_s = \omega_s + \widetilde{\text{$\Omega$}}_Q - \text{$\Omega$}\left( r_Q \right)$. Noting that
$\text{$\Omega$}\left( r_Q \right)$ is also a function of $\omega_s$, we may invert this relation to obtain
$\omega_s$ as a function of $\widetilde{\omega}_s$. A straightforward substitution for $\omega_s$ in terms of
$\widetilde{\omega}_s$ in Equation~(\ref{eq:BComega}) then yields Equation~(\ref{eq:finalomega}).

\begin{figure}
\begin{center}
\includegraphics[width=14.0cm,trim={0.0cm 8.2cm 0.5cm 9.0cm},clip]{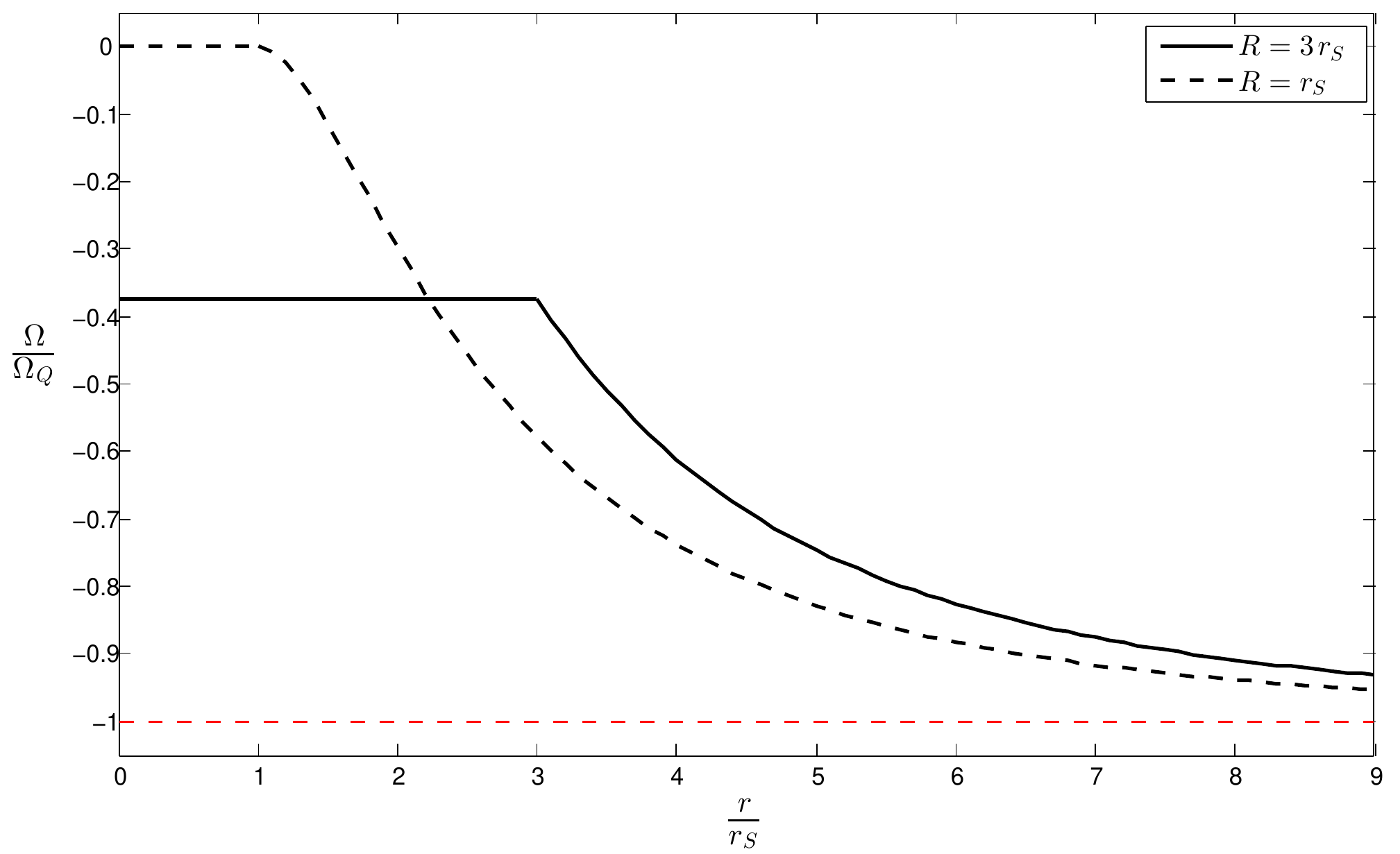}
%\vspace{-0.1cm}
\caption{Plots of the angular velocity $\text{$\Omega$}\left( r \right)$ of ZAMOs as given by the expression in
Equation~(\ref{eq:finalomega}) with vanishing angular velocity for the shell, $\omega_s=0$, and
$r_Q\rightarrow\infty$ as in Figure~\ref{fig:gensolinf}. Note that here
$\text{$\Omega$}\left( r \right)$ is normalized by $\text{$\Omega$}_Q$.}
\label{fig:solzeroshell}
\end{center}
\end{figure}

Finally, as a rather vivid illustration of the arbitrariness of the numerical value of $\text{$\Omega$}\left( r \right)$,
we may now consider the case, shown in Figure~\ref{fig:solzeroshell}, for which there is perfect dragging ($R=r_S$) and
the angular velocity of the shell vanish, $\omega_s=0$, and we keep $r_Q\rightarrow\infty$ as above
(but let $\text{$\Omega$}_Q$ be arbitrary).
Now both the shell and the inertial frames in the interior of the shell
are non--rotating, but the ZAMOs located at asymptotic infinity rotate with the angular velocity $\text{$\Omega$}_Q$.
In this picture it appears as if the massive shell and space in its interior are at rest while it is the rest of the
universe exterior to the shell which rotate around it.

\section{Rotation in the Kerr spacetime\label{sect:Kerr}}

In the previous section we have discussed rotation of ZAMOs in Brill and Cohen's approximate spacetime model of
a slowly rotating thin shell. In this section we briefly consider the angular velocity of ZAMOs in the Kerr spacetime
representing an exact axistationary and asymptotically flat solution to Einstein's field equations.

In Boyer--Lindquist coordinates $\left(t,r,\vartheta,\varphi\right)$ the angular velocity of ZAMOs due to inertial
dragging in the equatorial plane of the Kerr spacetime is given by~\cite{Gron1}
\begin{equation}
\label{eq:Kerrzamo}
\text{$\Omega$}\left( r \right) = \frac{2Ma}{r^3+ra^2+2Ma^2}\,,
\end{equation}
where $a=J/M$, $J$ is the angular momentum and $M$ is the mass characterizing the Kerr geometry. In the asymptotic
limit $r\rightarrow\infty$, $\text{$\Omega$}\left( r \right)$ vanishes. However, this does \emph{not} imply that the
ZAMOs located at asymptotic infinity define an absolute state of non--rotation. Indeed, using the same transformations
as in Equation~(\ref{eq:transf}), we may once again transform to a coordinate system
$\left({\widetilde{t},\widetilde{r},\widetilde{\vartheta},\widetilde{{\varphi}}}\right)$ rotating relative to the
Boyer--Lindquist coordinates to obtain the angular velocity of the ZAMOs in the rotating system as
\begin{equation}
\label{eq:Kerrzamorot}
\widetilde{\text{$\Omega$}}\left( r \right) = \widetilde{\text{$\Omega$}}_Q
+ \text{$\Omega$}\left( r \right) - \text{$\Omega$}\left( r_Q \right)\,.
\end{equation}
The ZAMOs are now seen to rotate at asymptotic infinity with, in general, a non--zero angular velocity
$\widetilde{\text{$\Omega$}}_Q - \text{$\Omega$}\left( r_Q \right)$ even though their angular momentum is zero.
If in addition we let the arbitrary reference radius $r_Q\rightarrow\infty$, then their rotation rate at infinity
equals the arbitrary reference value $\widetilde{\text{$\Omega$}}_Q$.

This observation may at first appear completely trivial since it follows directly from a simple transformation to
a rotating coordinate system. However, as was discussed below Equations~(\ref{eq:finalomega}) and (\ref{eq:BComega}),
this particular freedom of choice of coordinate system was inherently linked to the observation that Einstein's equations
determine only differences in angular velocities of ZAMOs, implying that only relative angular velocities are
meaningful concepts. The angular velocity (\ref{eq:Kerrzamo}) obtained in Boyer--Lindquist coordinates appears to be a
consequence of imposing the asymptotic boundary condition
$\displaystyle \lim_{r\rightarrow \infty} \text{$\Omega$}\left( r \right) = 0$ at the outset of the
derivation~\cite{Chandrasekhar}, potentially leading to the misconception that the inertial frames at asymptotic infinity
single out a global standard of non--rotation. The coordinate independent boundary condition of asymptotic flatness is
independent of the asymptotic boundary condition imposed on the numerical value of the ZAMO angular velocity. This
indicates that rotation of ZAMOs located at different radii is best interpreted as a relative concept in the
Kerr spacetime too.

\section{Summary and conclusion\label{sect:conclusion}}

The effect of inertial frame dragging on the rotation rate of ZAMOs has been analyzed within the framework of the simple
thin shell model seminally introduced by Brill and Cohen. By relaxing the quite natural assumption of zero rotation
infinitely far away from the massive shell early on in the derivation, the obtained expression for the rotation rate,
Equation~(\ref{eq:finalomega}), makes it clear that Einstein's equations only determine relative angular velocities of
ZAMOs. The particular value of the rotation rate of a ZAMO is physically irrelevant unless one also specifies an
arbitrarily chosen zero point in space relative to which this rotation rate is measured. Notably this applies as well
to the rotation rate of ZAMOs located at asymptotic infinity.

Within the simple thin shell model it was further clarified that the same expression for the rotation rate can be
obtained simply by a transformation from the coordinate system in which the ZAMOs at asymptotic infinity are
non--rotating to a coordinate system rotating relative to the first one. Utilizing this connection we then argued
that ``global'' rotation of ZAMOs in Kerr spacetime should be interpreted as a relative concept in the same sense
as for the thin shell model.

Thus, if there is still some hope that general relativity can be considered a complete (classical) theory of
gravitation capable of describing isolated gravitating systems without requiring particular topologies or
external causes such as ``fixed stars'' and absolutely non--rotating Lorentz frames at infinity, then it seems that
relative rotation in the sense presented here not only is a valid concept, but perhaps even a necessary one in the
interpretation of some solutions to Einstein's equations; more specifically, in the axistationary asymptotically
flat spacetimes of general relativity.

The question of whether rotational motion can be interpreted as relative according to general relativity has been
discussed in a relatively recent paper by \O.~Gr\o n~\cite{Gron2}. The arguments presented above provide some support
to this possibility. It seems feasible that both the absolute and relational viewpoints can be treated as
complementary aspects within the theory of general relativity~\cite{Overduin1,Overduin2}.

\vspace{6pt}
\small{\textbf{Acknowledgments:} The author would like to thank \O yvind Gr\o n for valuable suggestions and comments.}

\small{\textbf{Conflicts of Interest:} The author declares no conflicts of interest.}

\end{document}